\RequirePackage{amsmath}
\documentclass{svproc}
\usepackage{graphicx}
\usepackage[caption=false]{subfig}

\DeclareMathOperator{\arccosh}{arccosh}

\usepackage{url}

\begin{document}
\mainmatter              
\title{Latent space generative model for bipartite networks}
\titlerunning{Bipartite generative model}  
%
\author{Demival Vasques Filho\inst{1} $^{orcid.org/0000-0002-4552-0427}$ \and Dion R. J. O'Neale\inst{2}}
\authorrunning{Vasques Filho and O'Neale} 
%
%
\institute{Leibniz-Institut f\"ur Europ\"aische Geschichte \\ Alte Universit\"atsstra{\ss}e 19, 55116 Mainz, Germany\\
\email{vasquesfilho@ieg-mainz.de},\\ WWW home page:
\texttt{https://www.ieg-mainz.de/en/institute/people/\homedir vasques\_filho}
\and
Te P\={u}naha Matatini, Department of Physics, University of Auckland \\ Private Bag 92019, Auckland, New Zealand}

\maketitle              

\begin{abstract}
Generative network models are extremely useful for understating the mechanisms that operate in network formation and are widely used across several areas of knowledge. However, when it comes to bipartite networks --- a class of network frequently encountered in social systems --- generative models are practically non-existent. Here, we propose a latent space generative model for bipartite networks growing in a hyperbolic plan. It is an extension of a model previously proposed for one-mode networks, based on a maximum entropy approach. We show that, by reproducing bipartite structural properties, such as degree distributions and small cycles, bipartite networks can be better modelled and one-mode projected network properties can be naturally assessed. 
\end{abstract}
\section{\label{sec:introduction}Introduction}
Generative models are a powerful approach to describe and understand the processes at work during network formation and the mechanisms producing specific network features. They provide the opportunity to simulate real, growing networks, subject to various assumptions about the importance of controlled parameters \cite{orsini2015quantifying,denny2016importance}. Properties like heterogeneous degree distributions, clustering and community formation in real-world systems can be assessed using such models. 

Generative models have been developed in many flavours, associated with the different communities using them \cite{goldenberg2010survey,jacobs2014unified}. However, the same is not true for generative models for bipartite networks. Studies on statistical models for networks with bipartite structure are rare and even scarcer on generative models. Furthermore, the few studies addressing models of bipartite networks mostly focus on mimicking the properties of their projections only. Structural properties of bipartite networks are generally neglected.

We propose a latent space model in a hyperbolic plane, based on a maximum entropy approach, as an extension of work done specifically for one-mode networks \cite{krioukov2009curvature,krioukov2010hyperbolic,papadopoulos2012popularity}. We focus on recreating structural properties of bipartite networks, namely degree distribution and small cycles. The latter, especially four and six-cycles, have a significant effect on the resulting structure of the projected network. Four-cycles are indicative of recurring interactions, affecting the link weight distribution. Six-cycles, in turn, represent triadic closure and have an impact on the projected clustering \cite{vasques2019bipartite}. We show that, by reproducing such properties, the generative model produces bipartite networks whose one-mode projections naturally display the structures of interest.

The remainder of this paper is organised as follows. In Section \ref{null_models}, we examine the adaptation of null models for one-mode networks to generative models of bipartite networks. We discuss the characteristics of such models and how they fail to reproduce the main structural properties we are looking for in bipartite graphs. In Section \ref{sec:hyperbolic}, we discuss the popularity vs. similarity model \cite{papadopoulos2012popularity} for one-mode networks growing in a hyperbolic plane, based on a maximum entropy approach. In Section \ref{sec:generative}, we introduce our bipartite model and show how it recreates the features of real-world bipartite networks. Finally, we present the main results and the conclusion of the paper in Section \ref{sec:conclusion5}.

\section{\label{null_models}Null models} 

\subsection{Erd\H{o}s--R\'enyi}

The original Erd\H{o}s--R\'enyi model (ER) \cite{erdds1959random,erdos1960evolution} considers an ensemble of graphs $\mathcal{G}$, in which every graph $G \in \mathcal{G}$ has a set of nodes $U$, and $|L|$ links that connect pairs of nodes at random in the network. In a dynamic version of the model, we can add a node to the network at every time step $t$, until $|U|$ nodes are present. The number of links $|L|$, in turn, is controlled by adding $m$ new links to the network for every $t$, i.e. $|L|=tm$. Each graph has $m$ nodes at $t=0$ and at each time step a new node with $m$ links is added to the network, being randomly connected to $m$ existing nodes, until $t=|U|-m$.

\begin{figure*}[!ht]
	\centering
	\includegraphics[scale=0.35]{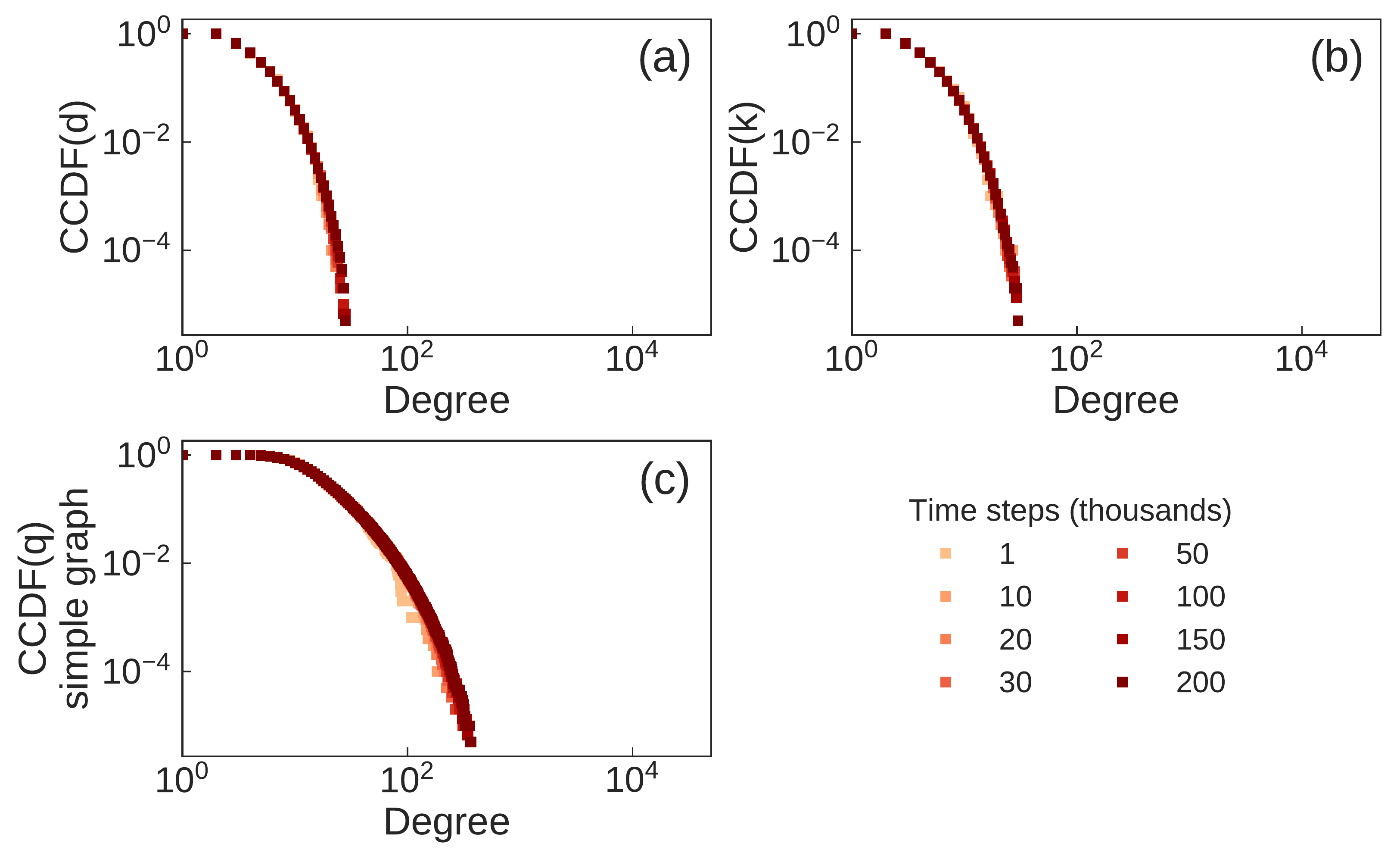}
	\caption{Degree distributions for the $\textrm{B}_{\textrm{ER}}(|U|,|V|,|E|)$ model. (a) Top and (b) bottom degree distributions are peaked and very similar, due to the mechanism of link attachment for both sets of nodes. (c) Projected networks follow the same degree distribution shape as the bottom node degree distribution, shifted to the right, as shown in \cite{vasques2018degree}.}
	\label{fig:erb_degree}
\end{figure*}

Based on this reasoning, we create a dynamic bipartite $\textrm{B}_{\textrm{ER}}(|U|,|V|,|E|)$ version of the ER model, where $|U|$ and $|V|$ are the number of bottom and tops nodes, respectively, and $|E|$ is the number of bipartite links, as follows:
\vspace{-0.25cm}
\begin{enumerate}
	\item At time $t=0$, the network has $m$ bottom nodes and $m$ top nodes, without links connecting them.
	\item At each time step, a new bottom node and a new top node are added to the network. The new top node chooses, at random, $m$ existing bottom nodes and connects to them. Then, for simplicity, this same process applies to the new bottom node which, in turn, connects to $m$ existing top nodes.
	\item Step 2 is repeated until the network has $|U|$ and $|V|$ bottom and top nodes, respectively. 
\end{enumerate}
\vspace{-0.25cm}
We implemented the above algorithm and used it to generate synthetic bipartite networks built with 200,000 time steps and $m=2$, such that $\langle d \rangle = \langle k \rangle = 4$. Due to the generative mechanism, the bottom and the top degree distributions are the same (Figure \ref{fig:erb_degree}). Moreover, the shape of the degree distribution of the projected network follows that of the bottom distribution, just shifted to the right, as shown in \cite{vasques2018degree}.

\begin{figure*}[!ht]
	\centering
	\subfloat{\label{fig:erb_cycles} \includegraphics[scale=0.25]{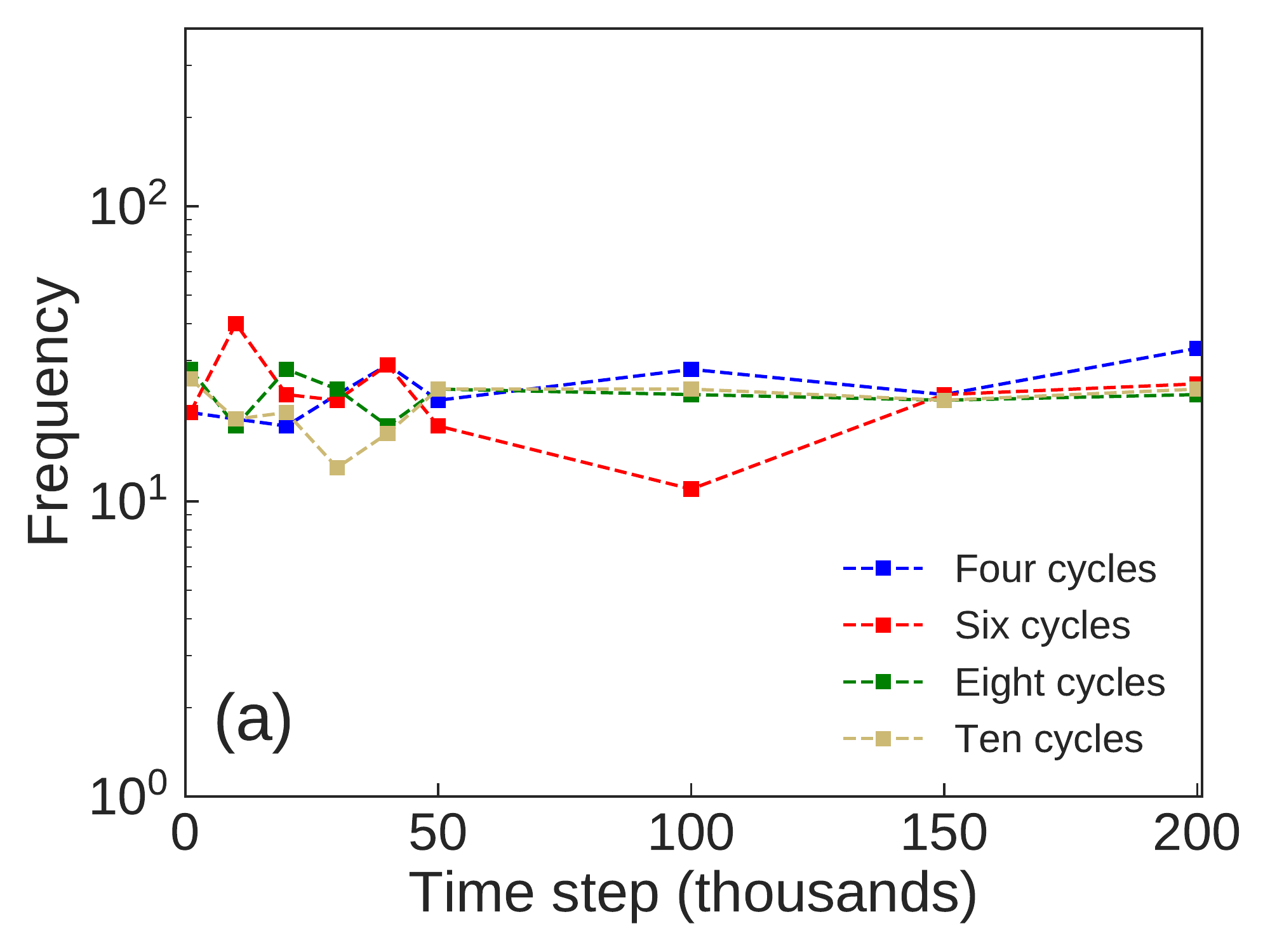}}%
	\subfloat{\label{fig:erb_links} \includegraphics[scale=0.25]{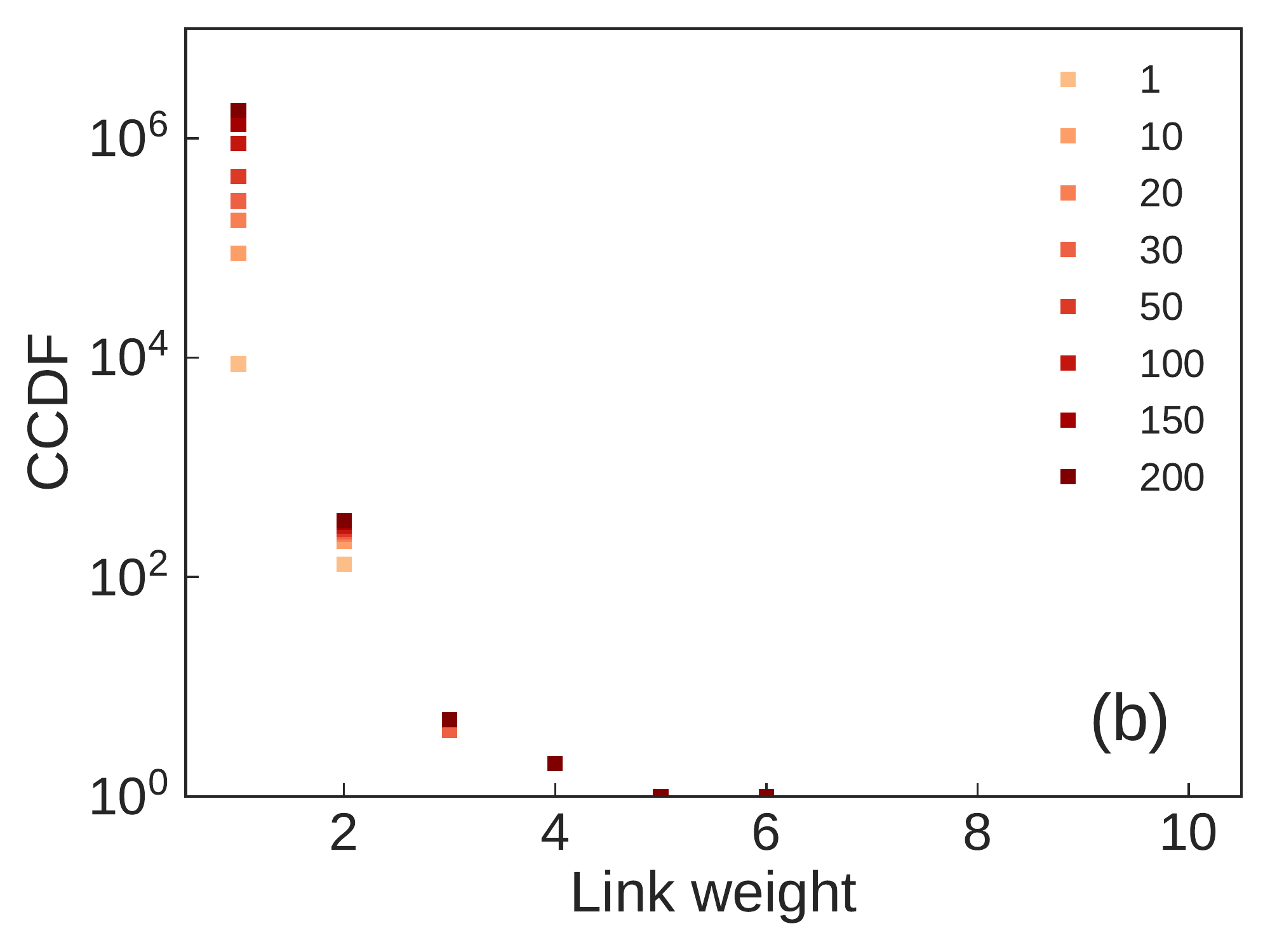}} \hfill
	\subfloat{\label{fig:erb_clustering} \includegraphics[scale=0.25]{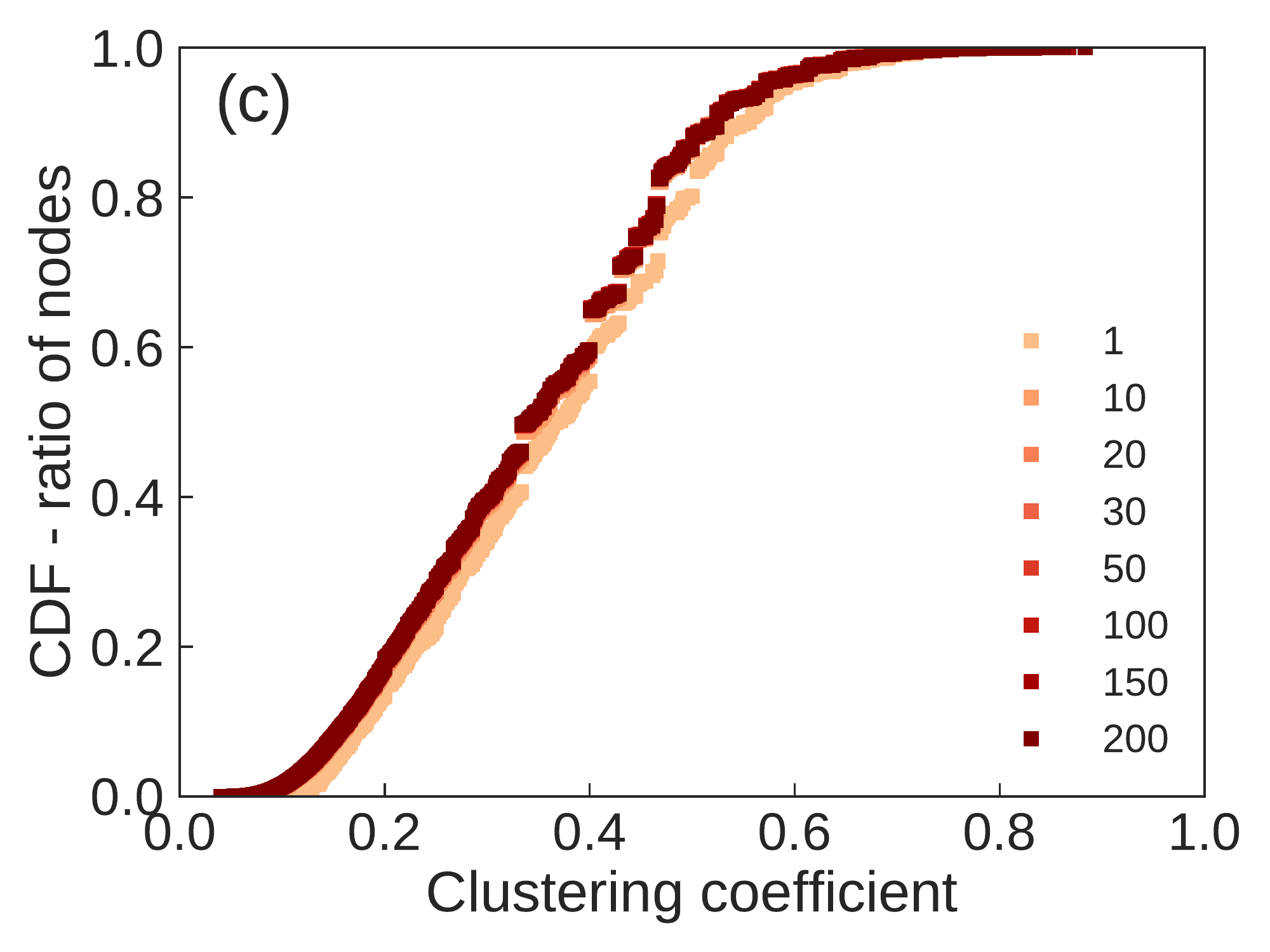}}%
	\caption{Evolution of (a) cycles in the bipartite network; and (b) link weight and (c) clustering distribution of the projected network for the $\textrm{B}_{\textrm{ER}}(|U|,|V|,|E|)$ model. As the network grows larger, the number of small cycles stays roughly constant, showing that random bipartite graphs tend to create uniform distributions of cycles size, as shown in \cite{vasques2019bipartite}. This results in few weighted links in the projected networks, and clustering created mostly by top node degrees instead of six-cycles.}
	\label{fig:erb_others}
\end{figure*}

The evolution of the number of small cycles (Figure \ref{fig:erb_cycles}) is roughly constant and at quite low levels, if compared to real-world networks \cite{vasques2019bipartite}. The same is true for the link weight (Figure \ref{fig:erb_links}) and clustering (Figure \ref{fig:erb_clustering}) distributions of the projected network. For the former, the absence of heavily weighted links is due to the low number of four-cycles, while for the latter, the low level of clustering is explained by the small number of six-cycles and by the absence of high-degree top nodes in the bipartite network. As expected, the generative version of the Erd\H{o}s--R\'enyi model still does not reproduce structural network properties of real networks. Let us explore next a preferential attachment bipartite generative model. 

\subsection{Preferential attachment}
Growing networks with preferential attachment has been extensively studied for both one-mode \cite{dorogovtsev2000structure,newman2001clustering,barabasi1999emergence} and bipartite networks \cite{peruani2007emergence,dahui2006bipartite,batagelj2005efficient,guillaume2006bipartite,chojnacki2012scale}. For the latter, our focus in this work, none of the models have addressed bipartite structural properties other than degree distributions. Furthermore, they have not investigated the effects of degree distributions on the one-mode projections.

Our preferential attachment generative model for bipartite networks is a bipartite version of the Barábasi-Albert (BA) model \cite{barabasi1999emergence}. It follows the same reasoning as the $B_{\textrm{ER}}$ generative model. The only difference is that new nodes now choose to connect to existing nodes from the opposite set with a weighted probability, where the weights are proportional to the degrees of the node in the target set. That is, 
\begin{equation}
\label{eq:bipartite_BA}
p_{u} = \frac{k_{u}}{\sum_{u'}k_{u'}}, \qquad p_{v} = \frac{d_{v}}{\sum_{v'}d_{v'}}\,,
\end{equation} 
where $k_{u}$ is the bottom degree of node $u$ and $d_{v}$ is the top degree of node $v$.

The model goes as follows:
\vspace{-0.25cm}
\begin{enumerate}
	\item At time $t=0$, the network has no links with only $m$ bottom and $m$ top nodes.
	\item At each time step, a new bottom node and a new top node enter the network. Now, the new top node chooses and connects to $m$ existing bottom nodes, with weighted probability according to Equation (\ref{eq:bipartite_BA}). Then, the new bottom node connects to $m$ existing top nodes, using the same formula for calculating the connection probability.
	\item Step 2 is repeated until the network reaches $|U|$ bottom and $|V|$ top nodes. 
\end{enumerate}

\begin{figure*}[!ht]
	\centering\includegraphics[scale=0.35]{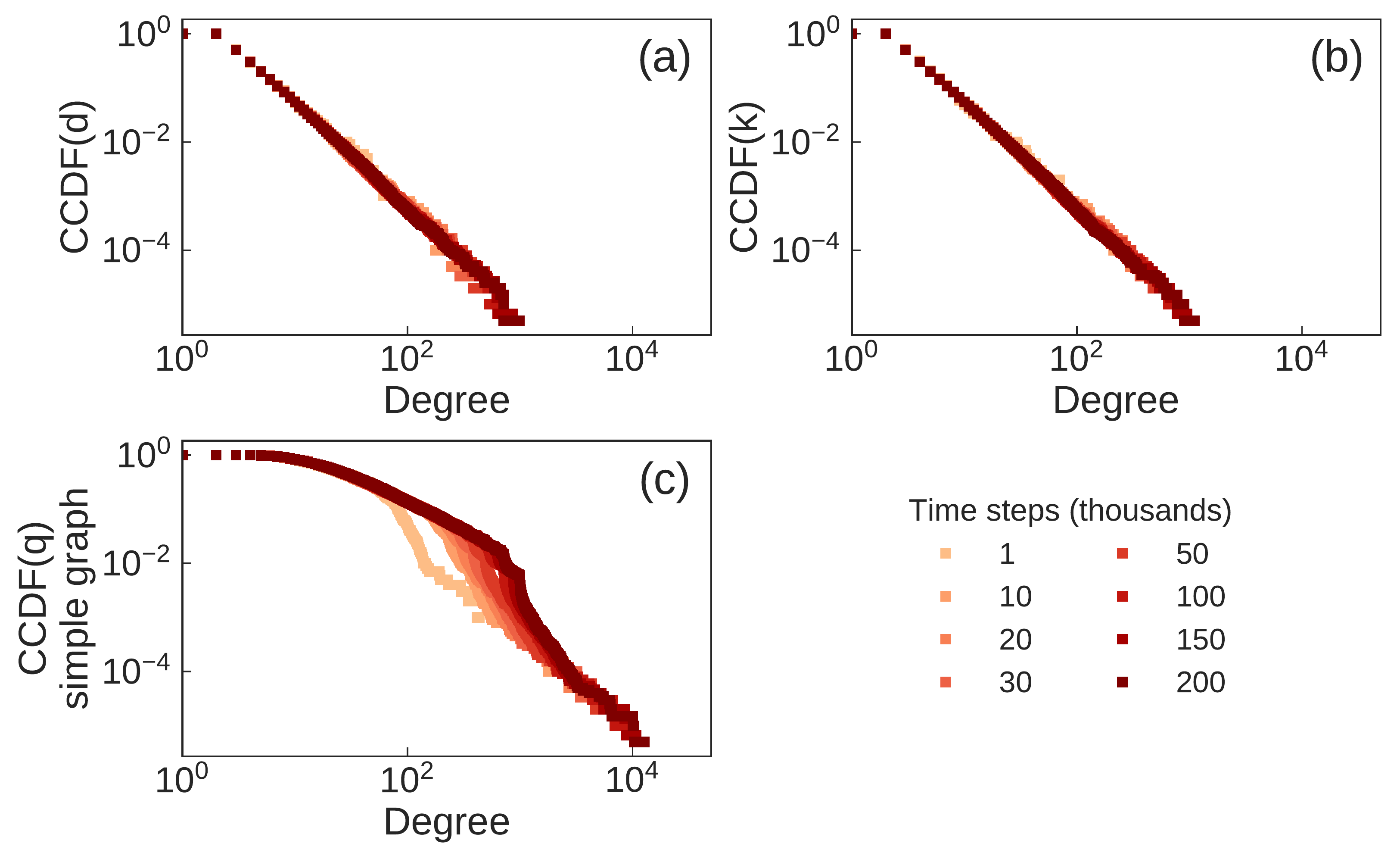}
	\caption{Degree distributions for the bipartite generative model of the Erd\H{o}s--R\'enyi model. Again, (a) top and (b) bottom degree distributions are very similar due to the mechanism of network growth. However the BA model creates heavy-tail degree distributions in this case. The behaviour of the (c) projected distribution deviates, especially for high-degree nodes, due to the cliques created in the projection.}
	\label{fig:bab_degree}
\end{figure*}

Again, we generate synthetic networks with $|U|=|V|=200,000$, with $m=2$, and $\langle d \rangle = \langle k \rangle = 4$. In Figure \ref{fig:bab_degree} we can see that, because of the simple preferential attachment mechanism of our model for both sets of nodes, the degree distributions for the top set of nodes, $P_{t}(d)$, and for the bottom nodes, $P_{b}(k)$, are the same. The degree distribution of the projected network, $P(q)$, is also heavy-tailed, but is shifted to the right, and shows a flattening similar to that shown in \cite{vasques2018degree}, due to the formation of large cliques, a consequence of the high-degree top nodes in $B$.

\begin{figure*}[!ht]
	\centering
	\subfloat{\label{fig:bab_cycles} \includegraphics[scale=0.25]{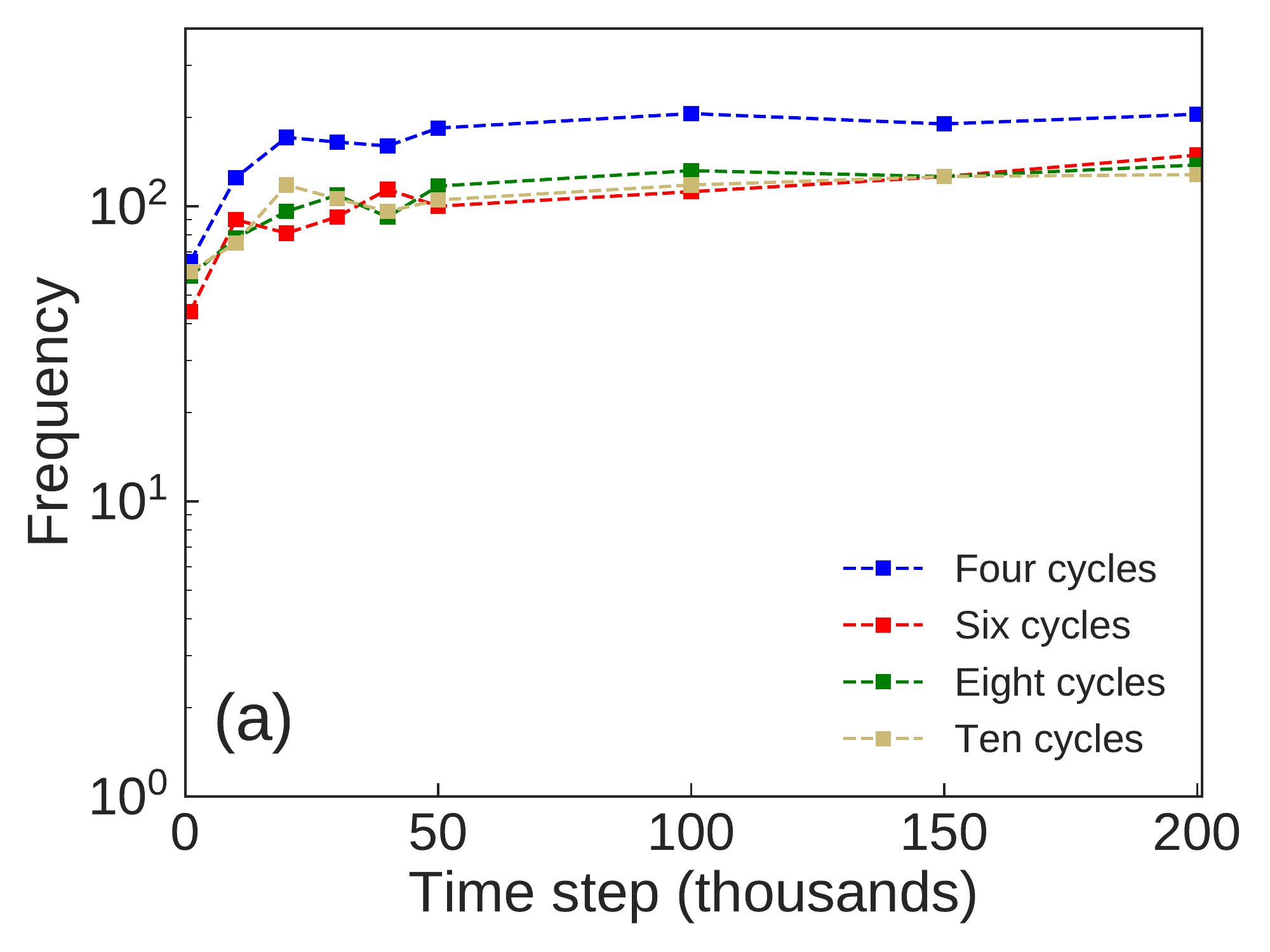}}%
	\subfloat{\label{fig:bab_links} \includegraphics[scale=0.25]{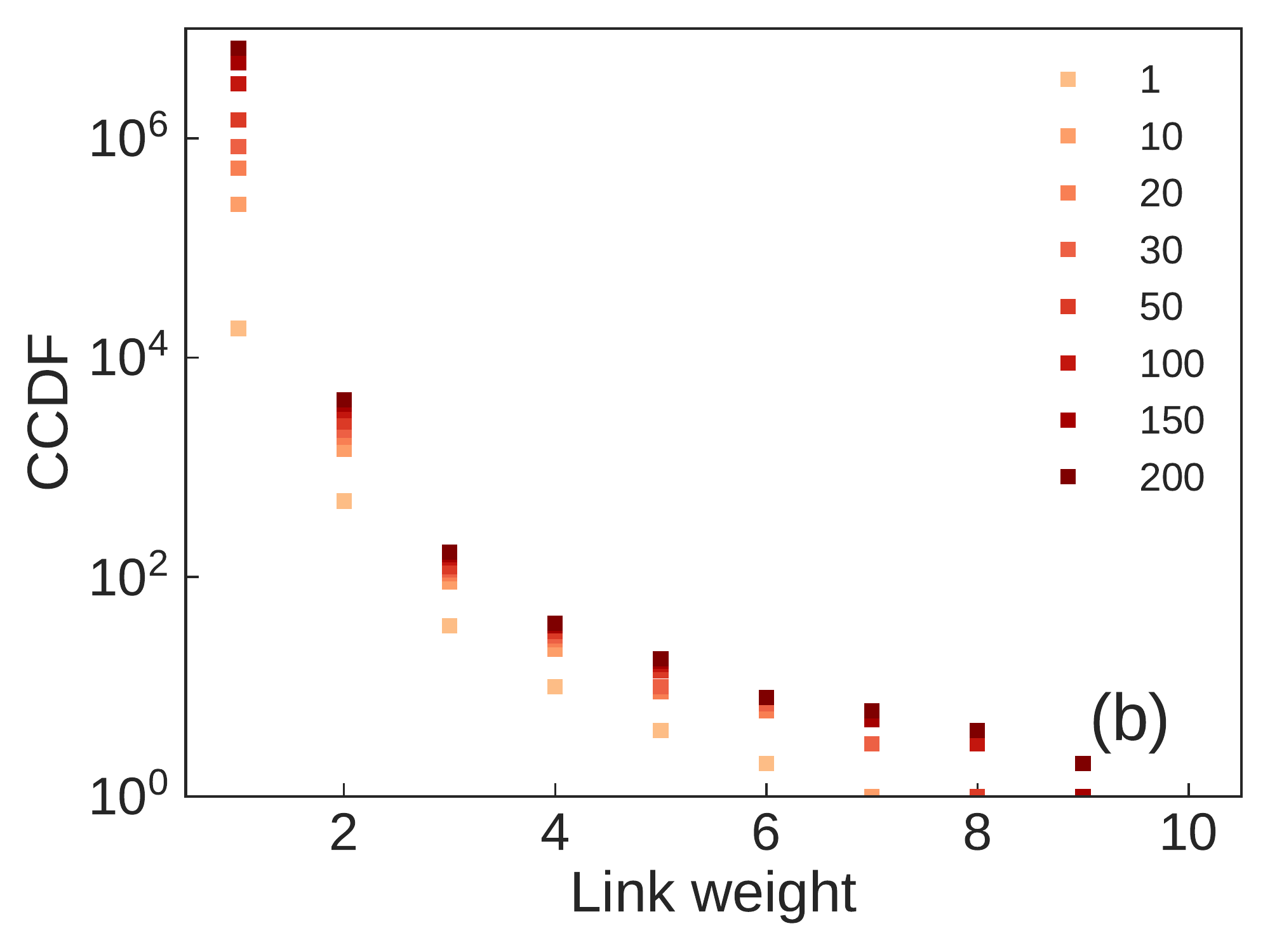}}\hfill
	\subfloat{\label{fig:bab_clustering} \includegraphics[scale=0.25]{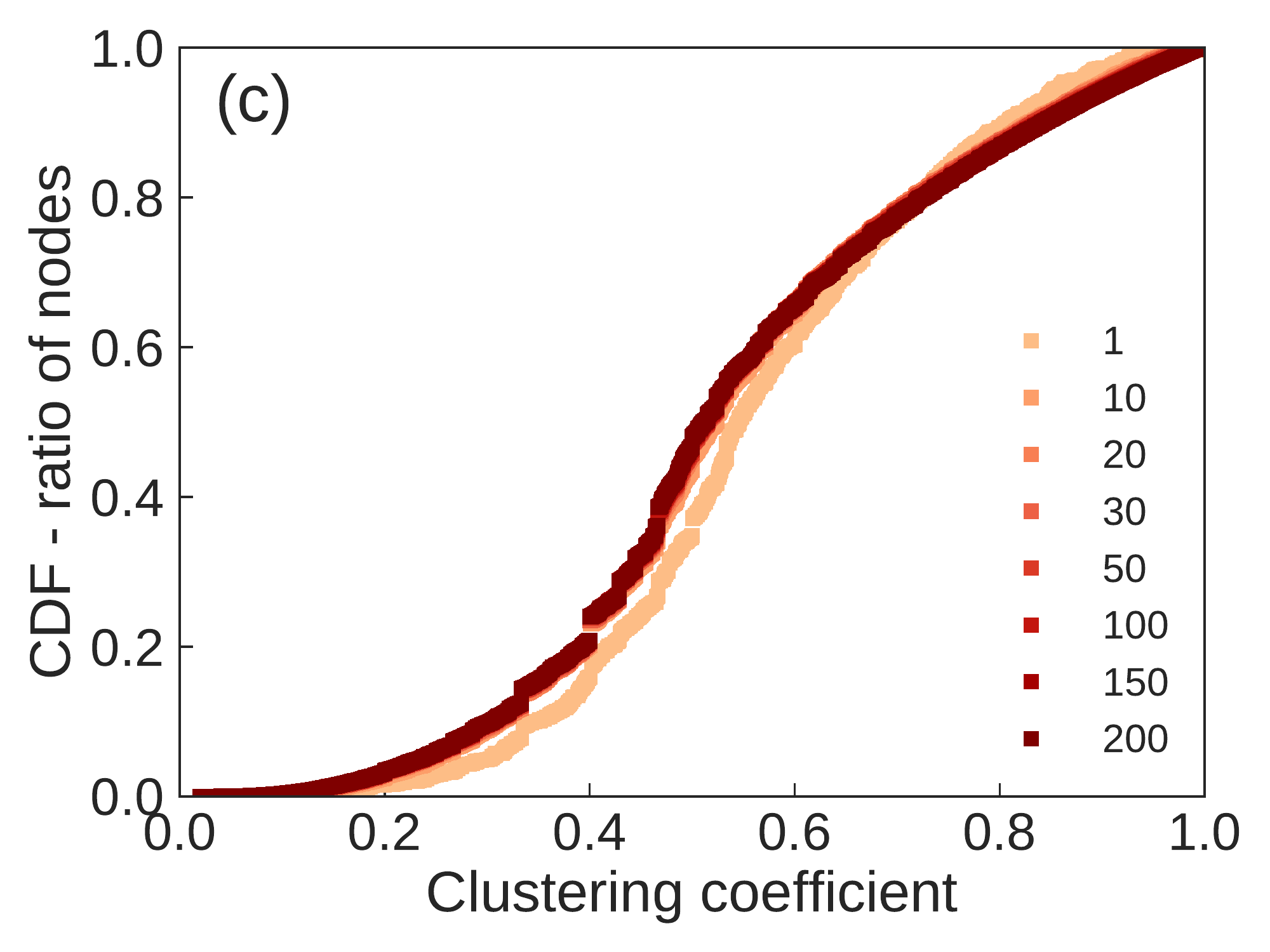}}%
	\caption{Evolution of (a) cycles in the bipartite network; and (b) link weight and (c) clustering distribution of the projected network for the bipartite BA model. Although the latter creates more small cycles than the $B_{\textrm{ER}}$ model, the presence of such cycles in the network is still very low compared to real-world networks. Still, it produces a few links with some weight and low levels of clustering in the projected networks.}
	\label{fig:bab_others}
\end{figure*}

The presence of high-degree nodes in the bipartite network increases, albeit only a little, the number of small cycles in the network (Figure \ref{fig:bab_cycles}). This is a result of a higher probability of high-degree top and bottom nodes being connected more frequently \cite{vasques2019bipartite}. However, the observed level of four-cycles is still relatively low compared to that seen in empirical networks and does not create a significant number of weighted links in the projected network $G_{w}$, as shown in Figure \ref{fig:bab_links}. Another consequence of the presence of high-degree top nodes in $B$ can be seen in Figure \ref{fig:bab_clustering}, where the level of clustering of the projection has increased relatively to the generative $B_{\textrm{ER}}(|U|,|V|,|E|)$ model. 

We have seen in \cite{vasques2019bipartite} that traditional null models as the $\textrm{B}_{\textrm{ER}}$ model and the configuration model --- both static --- cannot capture the structural properties of bipartite networks of our interest. Here, we have created synthetic networks with generative bipartite versions of the ER model and of the BA model. Although these dynamic models function well as null models, they do not reproduce the structure of real-world networks either. Hence, more sophisticated models are needed and we move in that direction in the next sections.

\section{\label{sec:hyperbolic}Hyperbolic geometry}
In a series of three papers \cite{krioukov2009curvature,krioukov2010hyperbolic,papadopoulos2012popularity}, it was demonstrated that some structural properties found in real-world networks, namely degree heterogeneity (heavy-tail degree distributions) and clustering, can emerge naturally when the network grows in a hyperbolic plane. The authors of \cite{papadopoulos2012popularity} used node coordinates in the hyperbolic plane as hidden variables \cite{binder1997adaptive,boguna2003class,serrano2008self,wu2012inferring}, characterizing their proposed popularity vs. similarity model as a latent space model.

However, that is not the only important characteristic of this model. The edge probability function chosen by the authors in \cite{krioukov2010hyperbolic} is the Fermi--Dirac distribution. The reason for that is threefold: first, the model incorporates the concepts of the exponential random graph models, through the maximum entropy approach \cite{park2004statistical}. Second, the model is initially designed for a simple graph one-mode network where links are fermions --- for a multigraph, or even a weighted network, the family of connection probabilities chosen would be the Bose--Einstein distribution, where more than one particle (link) could occupy the same energy state (pair of nodes) \cite{park2004statistical,garlaschelli2007interplay}. And third, because of the relation between statistical physics and the hyperbolic plane properties, as we will see next. 

In this model, the probability of two nodes being connected is, given by \cite{krioukov2010hyperbolic}
\begin{equation}
\label{eq:p_hyper}
P(u,u') = \frac{1}{e^{\beta \left(\frac{\zeta}{2}\right)(x_{u,u'}-R)}+1}\,,
\end{equation}
where, generalizing, we have
\begin{equation}
\omega = \frac{E - \mu}{kT} = \beta \left(\frac{\zeta}{2}\right)(x-R)\,.
\end{equation}
We can now interpret the set of auxiliary fields $\omega$. The hyperbolic distance $x$ between a pair of nodes in the network is the energy level occupied by the fermionic network links; $\zeta$ represents the curvature of the hyperbolic plane and plays the role of the Boltzmann constant; and the hyperbolic radius $R$ is the chemical potential. The inverse of temperature, $\beta$, acts as a input parameter, which can be used to control node coordinates and influence the strength of preferential attachment in the network, as we will see shortly.

The authors of \cite{papadopoulos2012popularity} proposed the one-mode generative model using the hyperbolic space with curvature $K=-\zeta^2=-4$ (so $\zeta=2$). In the simplest version of the model, each new node connects to the $m$ closest existing nodes, without the use of any connection probability function. However, we are interested in the more sophisticated --- grand canonical --- version of the model, where we have an expected number of links, instead of the exact number $|L|=mt$.

The model process works as follows \cite{papadopoulos2012popularity}:
\begin{enumerate}
	\item At time $t=0$, the network is empty.
	\item For every time step $t\geq1$, a new node enters the network with radial coordinate $r_{u} = \ln t_{u}$ and angular coordinate $\theta_{u}$ picked from a uniform random distribution on $(0,2\pi]$.
	\item Existing nodes $u'$, with $t_{u'}<t_{u}$, have their radial coordinates updated as 
	\begin{equation}
	\label{eq:update_r}
	r_{u'}(t) = \alpha r_{u'} + (1-\alpha)\ln t\,.
	\end{equation}
	The parameter $\alpha$ tunes the tail of the degree distribution. More specifically it gives the exponent of the power law, such that 
	\begin{equation}
	\label{eq:gamma}
	\gamma = 1+\frac{1}{\alpha}\,.
	\end{equation}
	That is, when $\alpha=1$ the radial coordinates are not being updated at all and we have a strong preferential attachment. On the other hand, when $\alpha \rightarrow 0$, all nodes move outwards from the center at the same speed, hence, we create a random network.
	\item The new node tries to connect to every existing node with probability given by Equation (\ref{eq:p_hyper}). The hyperbolic distance between a pair of nodes $u,u'$ is given by 
	\begin{equation}
	\label{eq:xuu'}
	x_{u,u'} = \frac{1}{2} \arccosh (\cosh 2r_{u'}\cosh 2r_{u} - \sinh 2r_{u'}\sinh 2r_{u} \cos \theta_{u,u'})\,,
	\end{equation} where $\theta_{u,u'} = \pi - |\pi - |\theta_{u'}-\theta_{u}||$.
\end{enumerate}
Let us take a closer look at the initial parameters of the model $m$, $T$. While $T$ appears in Equation (\ref{eq:p_hyper}), that is not the case for $m$. However, just like the other models, $m$ is a parameter that controls the number of links in the network. It affects the hyperbolic radius $R$, of Equation (\ref{eq:p_hyper}), at time $t$, according to \cite{papadopoulos2012popularity}
\begin{equation}
\label{eq:Rt}
R_{t} = \ln t - \ln \left[\frac{2T}{\sin T\pi}\frac{(1-e^{-(1-\alpha)\ln t})}{m(1-\alpha)}\right]\,,
\end{equation}
in such a way that the average degree of the network still follows $\langle q \rangle = 2m$.

Finally, the temperature $T$ of the system functions, in the model, as one might expect. As $T$ increases, higher energy levels can be occupied by our particles, and more disorder is observed in the system. For our model, this translates to having the probability of connection between distant nodes increasing with $T$ (Equation (\ref{eq:p_hyper})). Thus, temperature controls the level of clustering of the network. As $T \rightarrow 0$, we reach the strongest levels of clustering, as only nodes positioned closest to each other have high connection probabilities, creating triadic closure and, as a consequence, communities. $T$ takes values in the interval $(0,1]$, which is called the cold regime \cite{krioukov2010hyperbolic}. At values $T\geq1$ (hot regime) clustering levels are close to $0$, similar to those for the BA model.  

The hyperbolic model parameters generate popularity and similarity (hence the name given to the model), which are related to preferential linking and high clustering, respectively.

In summary, the radial coordinate $r$ and the parameter $\alpha$ determine the amount of preferential attachment in the network, while the angular coordinate $\theta$ and the parameter $T$ determine the strength of the clustering.
Papadopoulos \textit{et al.} provide elegant analytical solutions for the model, in \cite{papadopoulos2012popularity}, along with empirical validation for fitting the model to the Internet, the \textit{E. coli} metabolic network, and the PGP web of trust. However, they note that the model does not reproduce well the actor-movie network because of the over-inflation of connections --- the complete subgraphs --- created by the co-occurrence network they are considering. In other words, their proposed model fails to replicate a one-node projection of a bipartite social network. That is why we propose a bipartite version of the model in the next section.

\section{\label{sec:generative}Bipartite generative model}


We consider a bipartite generative model with two sets of nodes, $U$ and $V$, growing in the same plane, with the constraint that nodes of the same set cannot be connected in the bipartite network. We take the artifacts (top nodes) to be the nodes creating new links in the network, while the bottom nodes attract such links. In this way, artifacts only connect to agents in the time step when they enter the network. This process that we choose mimics, for instance, the processes of the scientific network, where papers do not gain links to additional authors after appearing in the network, but authors can continue to produce new publications (with potential co-authors) throughout their careers.

The model goes as follows:
\begin{enumerate}
	\item At time $t=0$, the network in empty.
	\item For every time step $t\geq1$, a new top node $v$ and a new bottom node $u$ enter the network with radial coordinates $r_{v} = \ln t_{v}$ and $r_{u} = \ln t_{u}$; and angular coordinates $\theta_{v}$ and $\theta_{u}$, drawn at random from a uniform distribution on $(0,2\pi]$.
	\item Existing bottom nodes update their radial coordinates according to Equation (\ref{eq:update_r}) (top nodes have fixed radial coordinates).
	\item The new top node $v$ connects to bottom nodes with a probability given by
	\begin{equation}
	\label{eq:puv}
	p_{(u,v)} = \frac{1}{e^{\frac{(x_{u,v}-R_{u})}{T}}+1}\,.
	\end{equation} 
\end{enumerate}

As before, the parameters of the model are $m$, $\alpha$ and $T$. Because top nodes are not attracting links, their degree distribution will always be roughly the same for every $\alpha$ and $T$. Moreover, the top degree distribution will always keep the same shape, but will be right-shifted as $m$ increases. 


There is, however, a way to relax the constraint of top nodes always having a very similar degree distribution. This can be done simply by drawing a value for $m$, in each time step, from a probability distribution. As $m$ changes, it affects the chemical potential of the system, given by Equation (\ref{eq:Rt}). Bigger values of $m$ result in higher $R$ which, in turn, increases the connection probability in the network. Hence, the shape of the tail of the top degree distribution can be easily tuned, while keeping the same $\langle d \rangle$. 

\begin{figure*}[!ht]
	\centering
	\subfloat{\label{fig:hgb_alpha1} \includegraphics[scale=0.25]{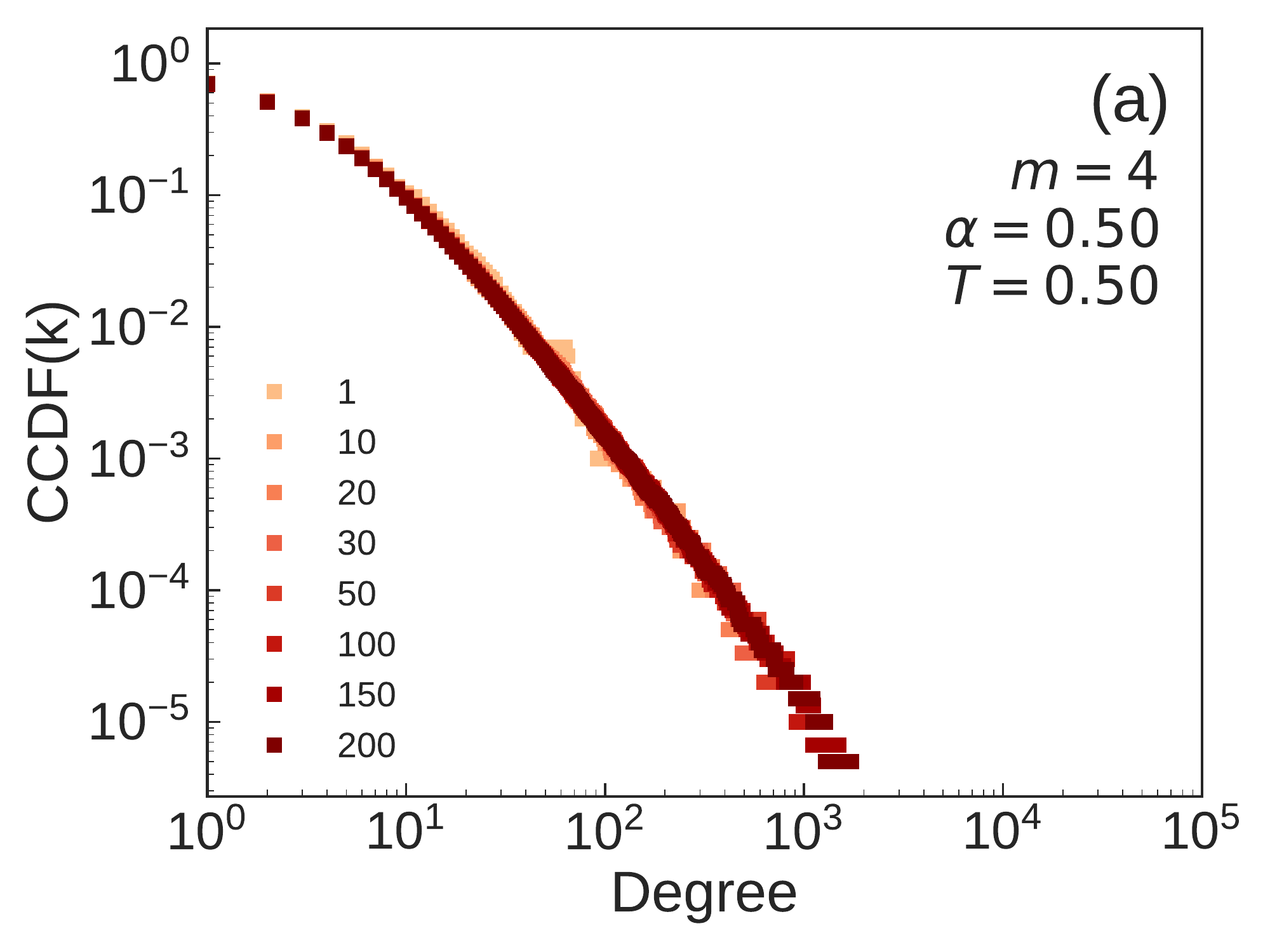}}%
	\subfloat{\label{fig:hgb_alpha2} \includegraphics[scale=0.25]{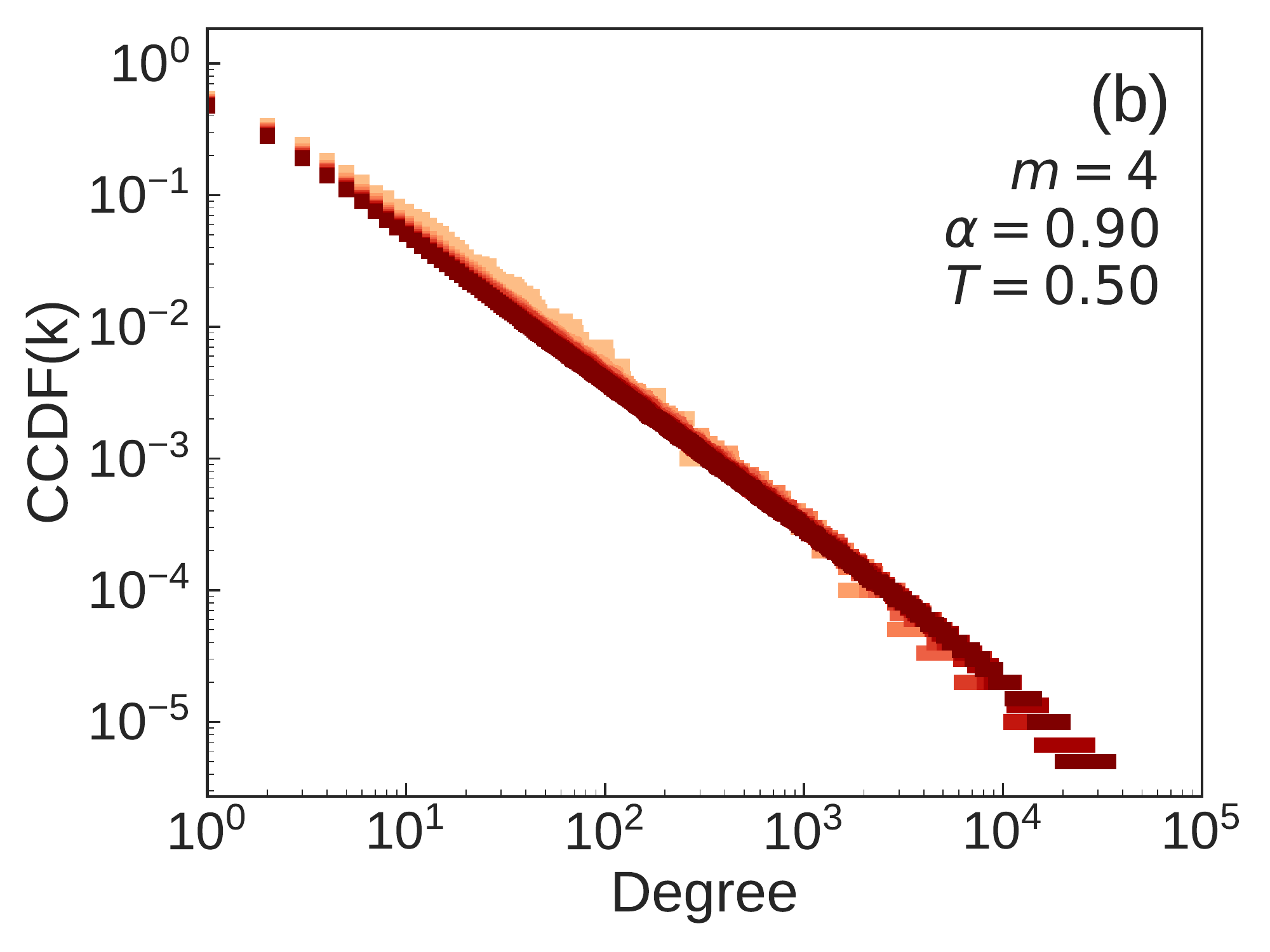}} \hfill
	\subfloat{\label{fig:hgb_alpha3} \includegraphics[scale=0.25]{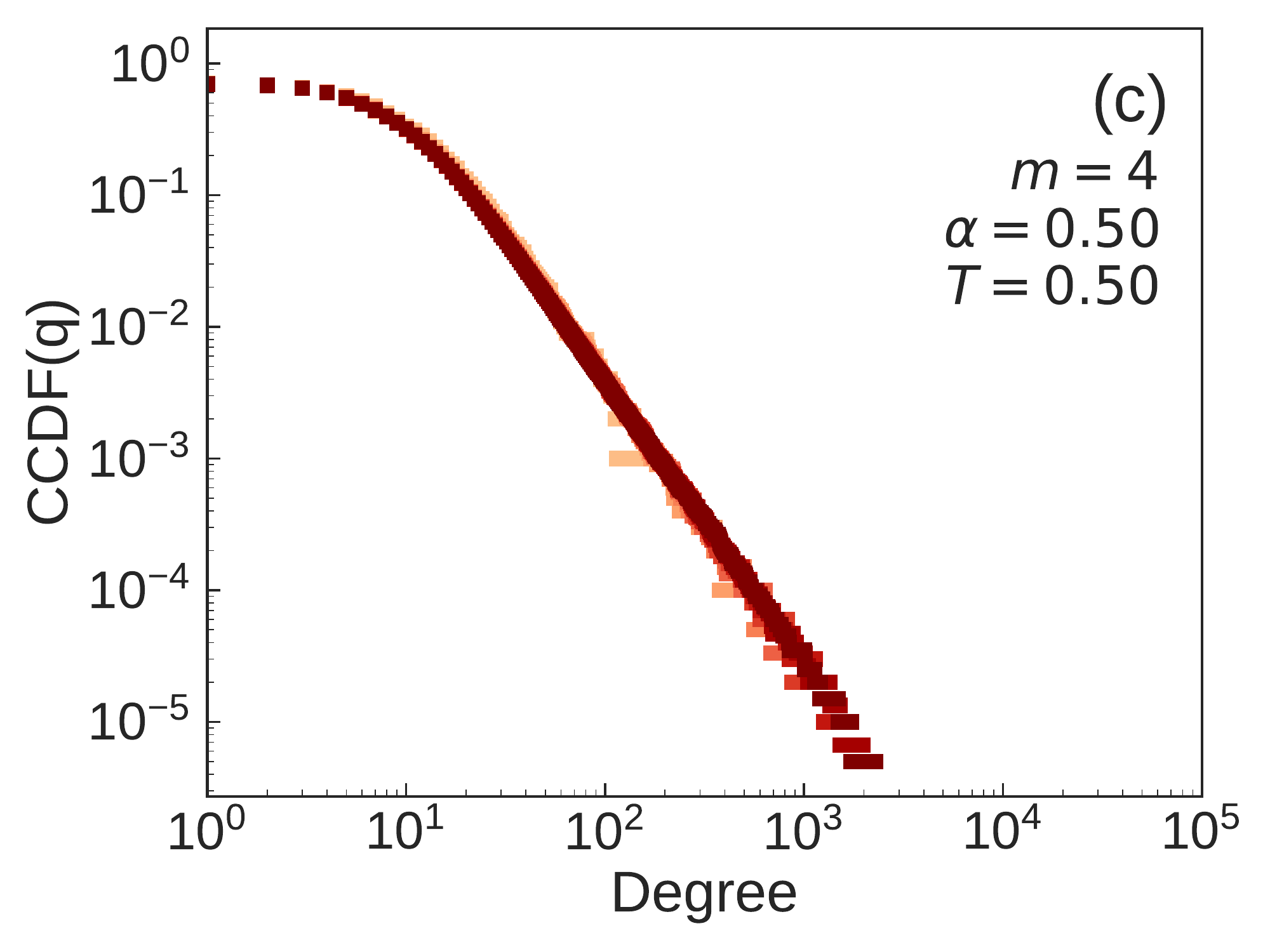}}%
	\subfloat{\label{fig:hgb_alpha4} \includegraphics[scale=0.25]{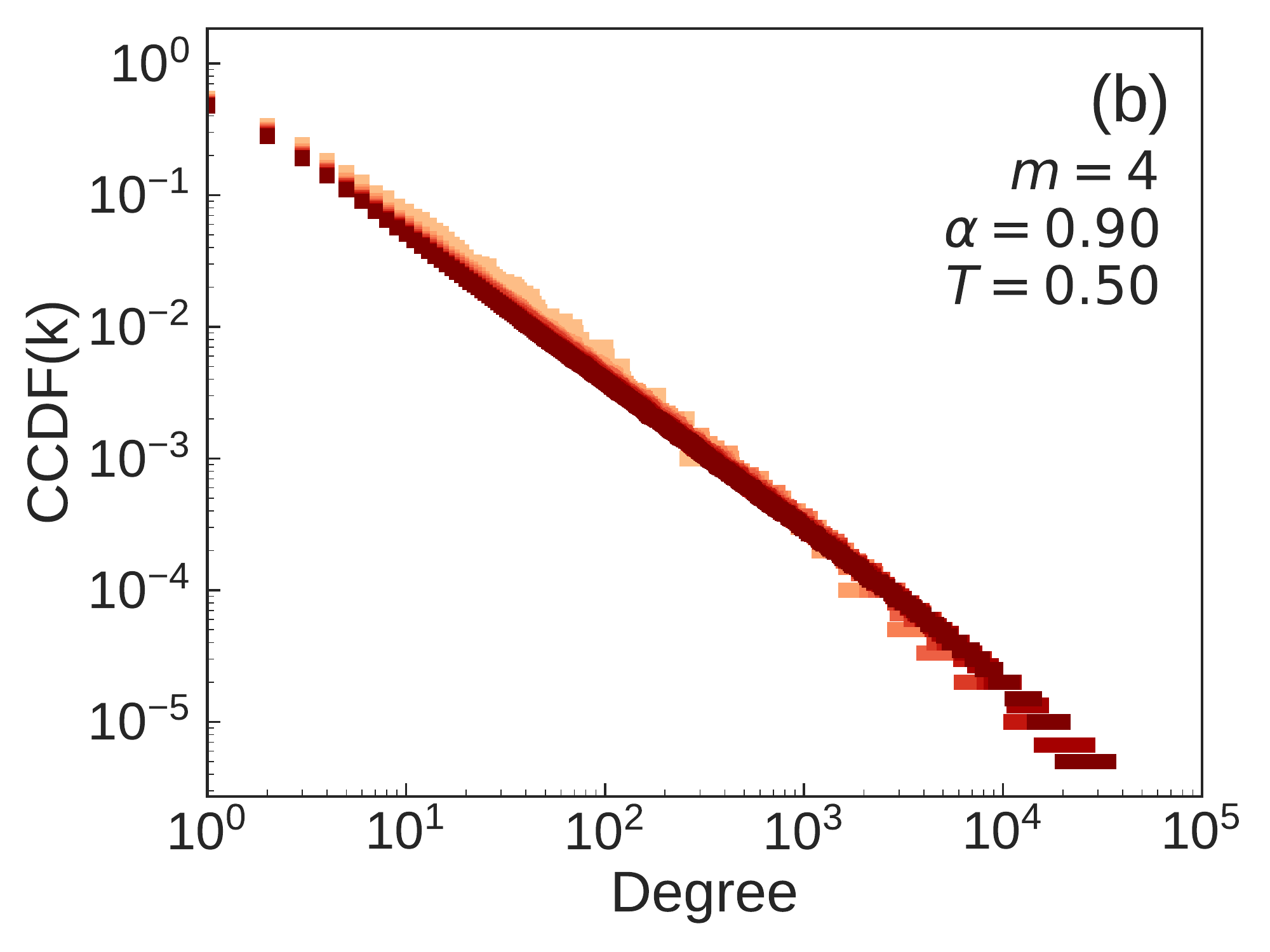}}%
	\caption{Bottom and projected degree distributions for synthetic networks built with variations of parameter $\alpha$. (a) $m=4$, $\alpha=0.50$ and $T=0.50$ for bottom and projected distributions, respectively; (b) change of parameter $\alpha=0.90$. (c) and (d) the same as (a) and (b), however for projected networks, respectively. We can clearly see the effect of the parameter $\alpha$ controlling the radial coordinates of the bottom nodes and, therefore, the level of preferential attachment in the network. Best fit for the bottom degree distributions gives us (a) $\gamma = 2.87$ and (b) $\gamma = 2.09$, compared to the predicted values $\gamma = 3$ and $\gamma = 2.10$ according to the analytical solution given by Equation (\ref{eq:gamma}). Moreover, the degree distributions of the projected networks are very similar to the bottom degree distributions. This is due to the fact that bottom distributions are more right-skewed than top distributions \cite{vasques2018degree}.}
	\label{fig:hgb_alpha}
\end{figure*}

The bottom degree distribution does not need any additional mechanism as the tail of the distribution is controlled by the parameter $\alpha$ (Figures \ref{fig:hgb_alpha1} and \ref{fig:hgb_alpha2}). Preferential attachment is guaranteed based on the time when nodes appear in the network.  This is explained by the radial coordinate, since early nodes are positioned closer to the origin of the hyperbolic plane, and therefore have a higher probability of being closer to more nodes in the network, according to Equation (\ref{eq:xuu'}). From Equation (\ref{eq:update_r}), we see that when $\alpha \rightarrow 1$ we have strong preferential attachment. As $\alpha$ decreases, so does the weight in the tail of the probability distribution, following Equation (\ref{eq:gamma}). On the other hand, for $\alpha \rightarrow 0$, the positions of every bottom node are updated, at each time step, moving to the edge of the hyperbolic disc (Figure 1c of \cite{papadopoulos2012popularity}). Hence, all nodes have the same connection probability, which characterises a random network regime for the bottom nodes. 

The shape of the projected degree distributions (Figures \ref{fig:hgb_alpha3} and \ref{fig:hgb_alpha4}) is in agreement with results shown in \cite{vasques2018degree}, following the bottom degree distribution. 

\begin{figure*}[!ht]
	\centering
	\subfloat{\label{fig:hgb_T1} \includegraphics[scale=0.25]{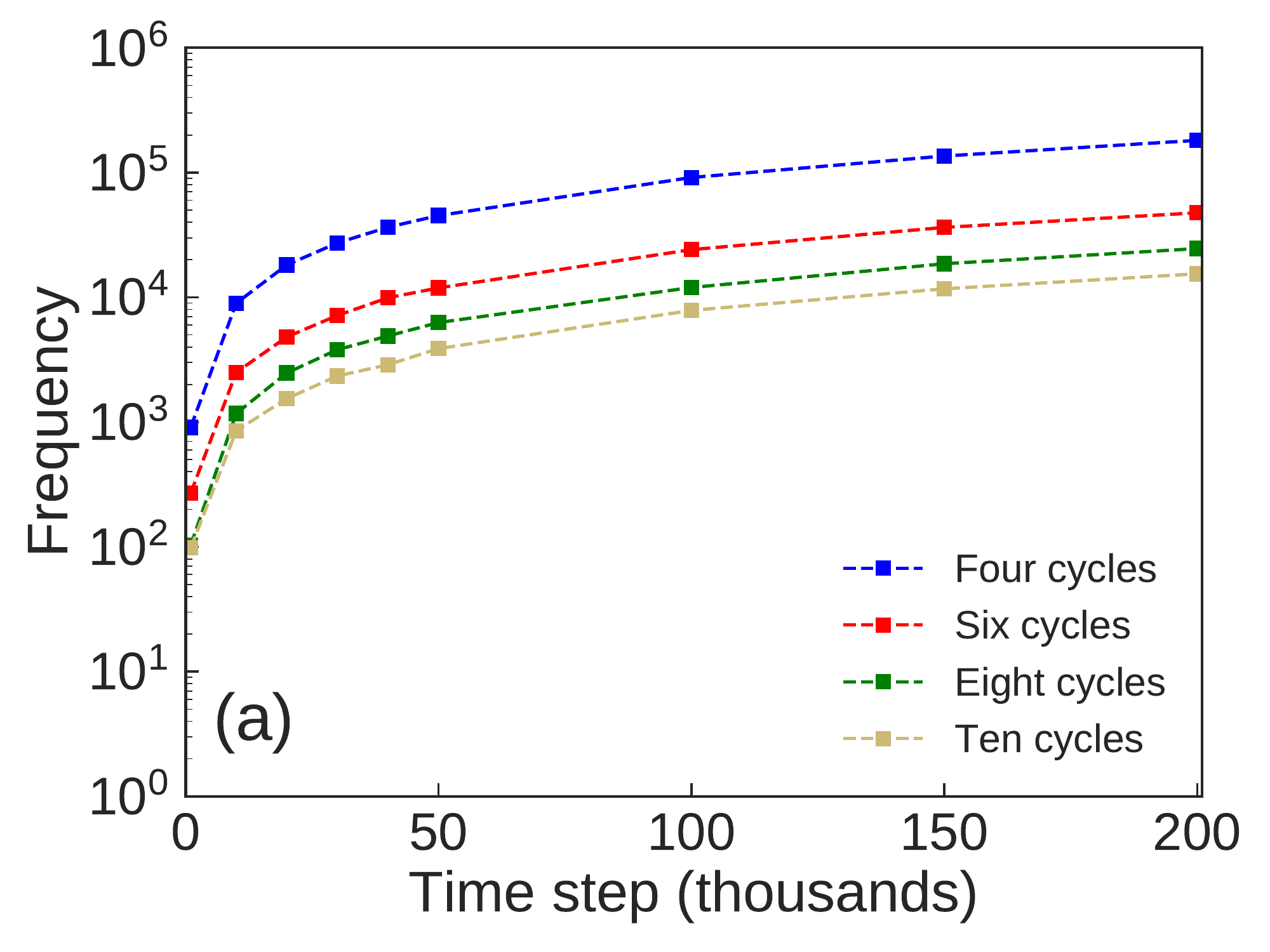}}%
	\subfloat{\label{fig:hgb_T2} \includegraphics[scale=0.25]{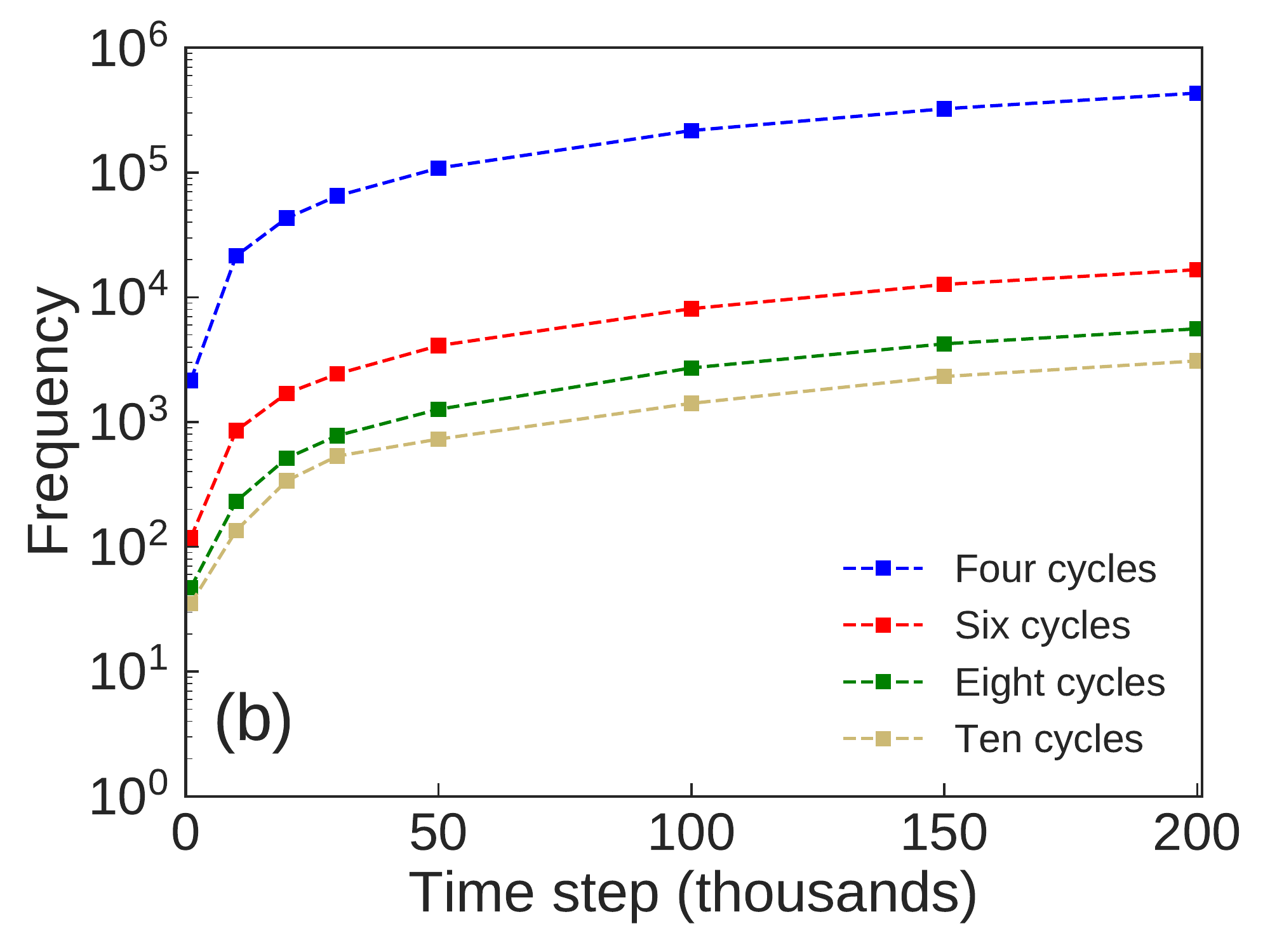}} \hfill
	\subfloat{\label{fig:hgb_T3} \includegraphics[scale=0.25]{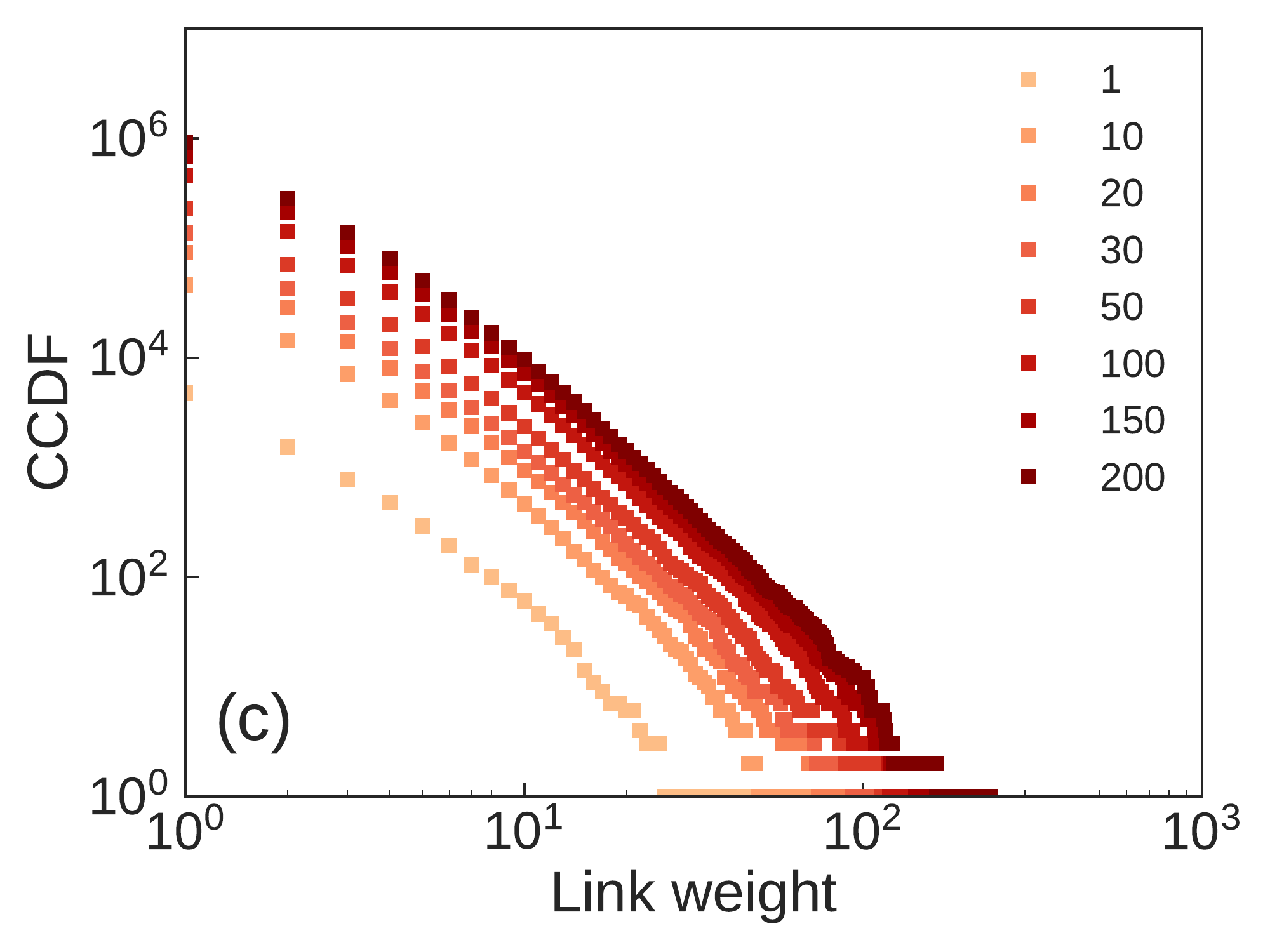}}%
	\subfloat{\label{fig:hgb_T4} \includegraphics[scale=0.25]{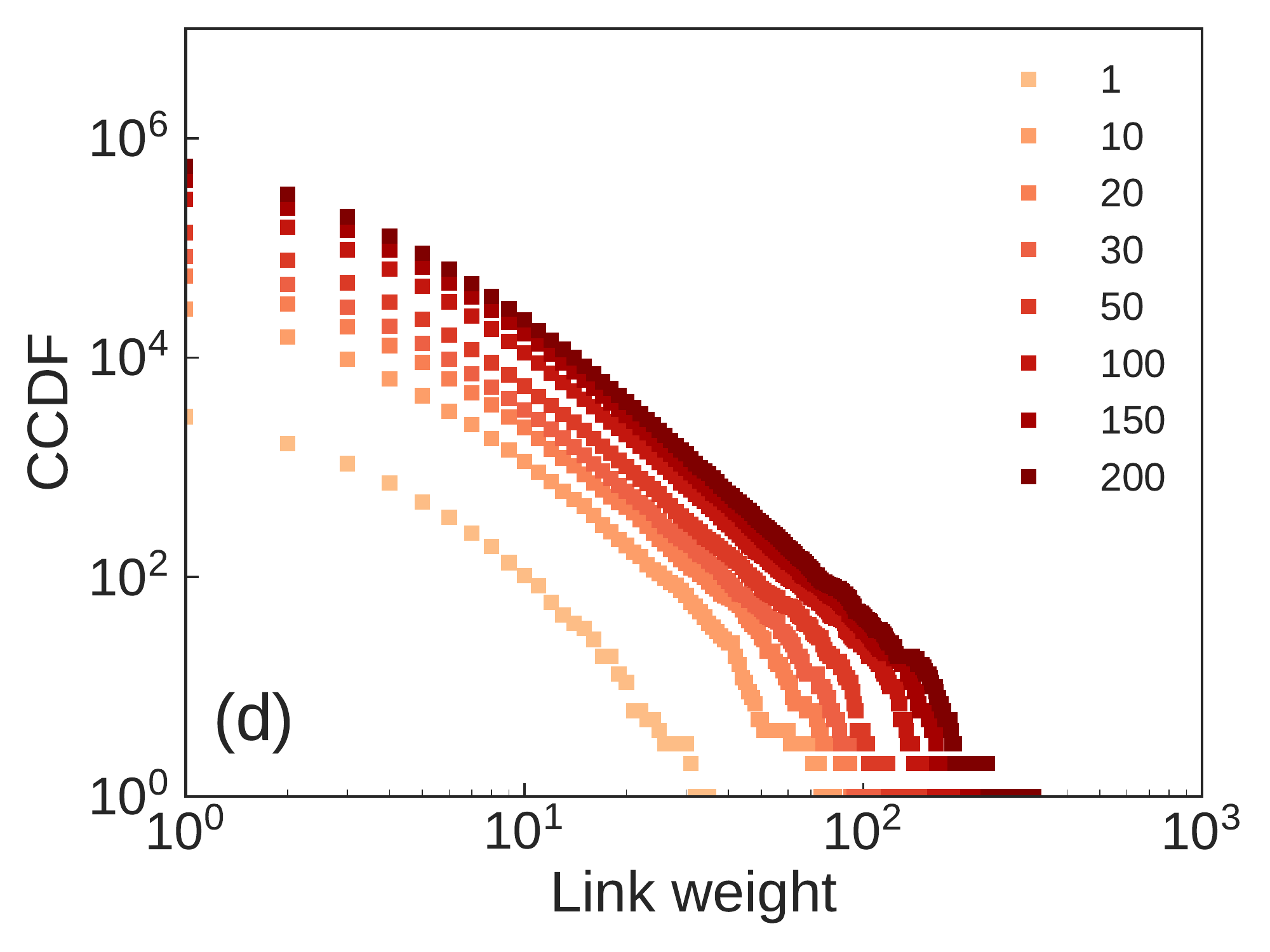}} \hfill
	\subfloat{\label{fig:hgb_T5} \includegraphics[scale=0.25]{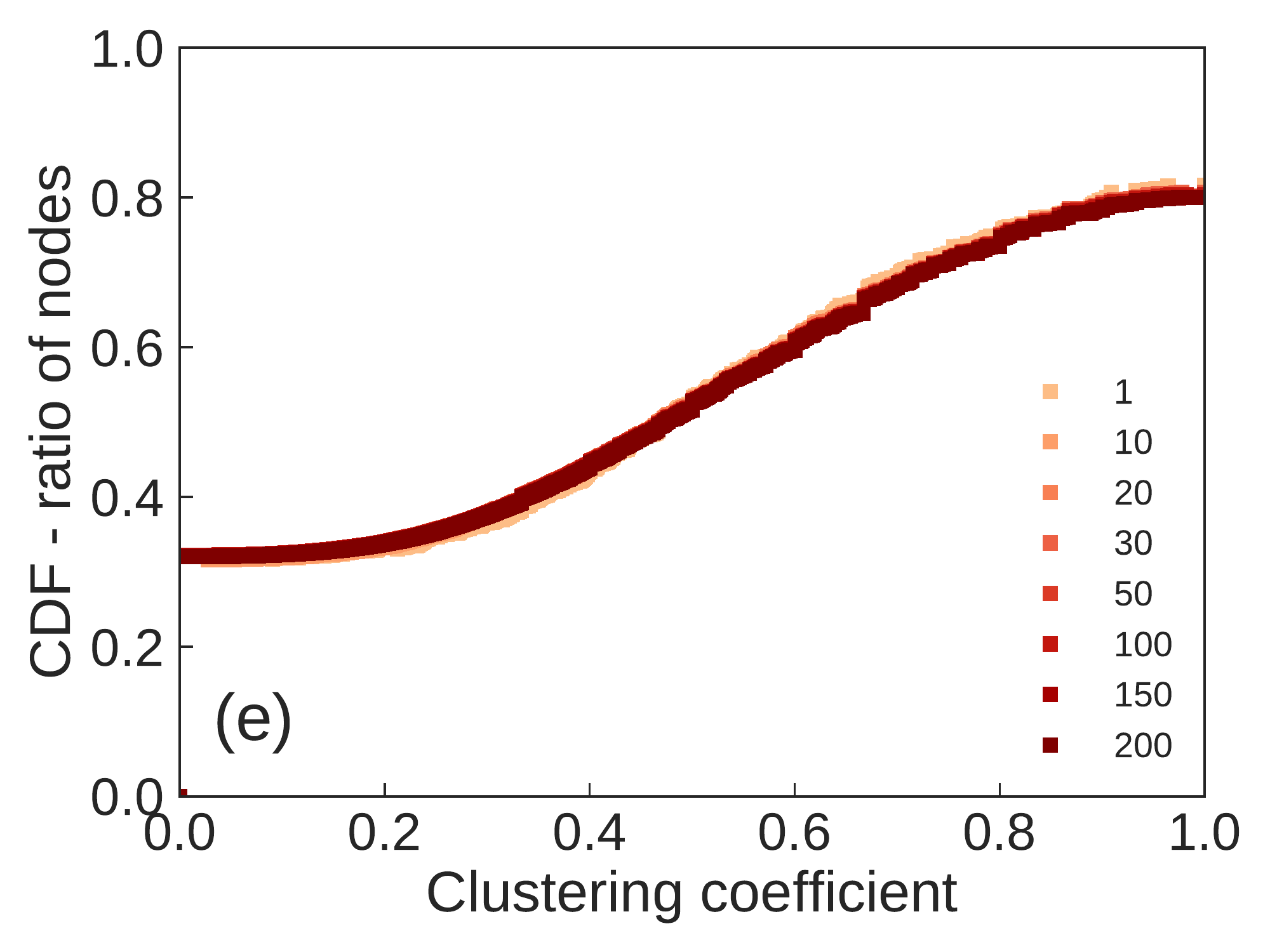}}%
	\subfloat{\label{fig:hgb_T6} \includegraphics[scale=0.25]{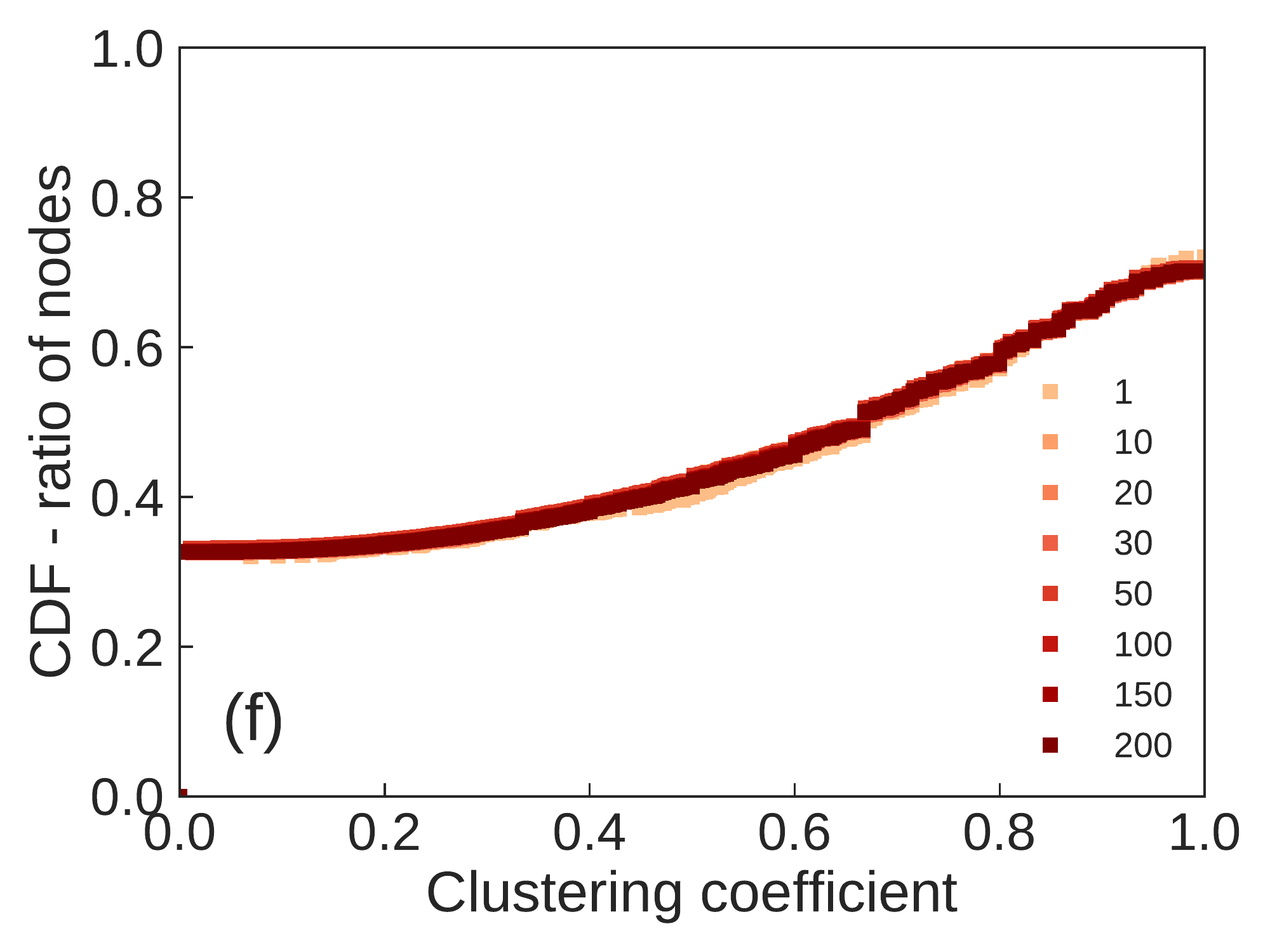}}
	\caption{(a) and (b) evolution of small fundamental cycles; (c) and (d) link weight distributions; (e) and (f) clustering distributions. Left column for network with temperature $T=0.50$ and right column with $T=0.10$. At higher temperatures the presence of four-cycles is smaller, and the gap between their frequency and the frequency of other cycles reduces. We can clearly see how more four-cycles in the network shift the link weight distribution. Note that the number of fundamental six-cycles decreases, yet the clustering coefficient increases. This is the effect of the substantial rise in the number of four-cycles, at low temperature, which hides larger cycles (Figure \ref{fig:4hiding6}), hence widening the gap between four-cycles and larger cycles.}
	\label{fig:hgb_T}
\end{figure*}

In order to control the number of small cycles present in the network, we use the last parameter of the model, $T$, the temperature of the system. Similarly to the case of one-mode networks, where $T$ tunes clustering (the number of triangles), for the bipartite version, $T$ primarily controls the presence of four-cycles, but also the presence of six-, eight- and 10-cycles. At lower temperatures, nodes that are closer in the plane have higher probabilities of being connected, favouring the presence of small cycles. We can see a substantial increase in the number of four-cycles in the network, widening the gap between them and the other small cycles, as shown in Figures \ref{fig:hgb_T1} and \ref{fig:hgb_T2}.    

\begin{figure*}[!ht]
	\centering
	\includegraphics[scale=0.60]{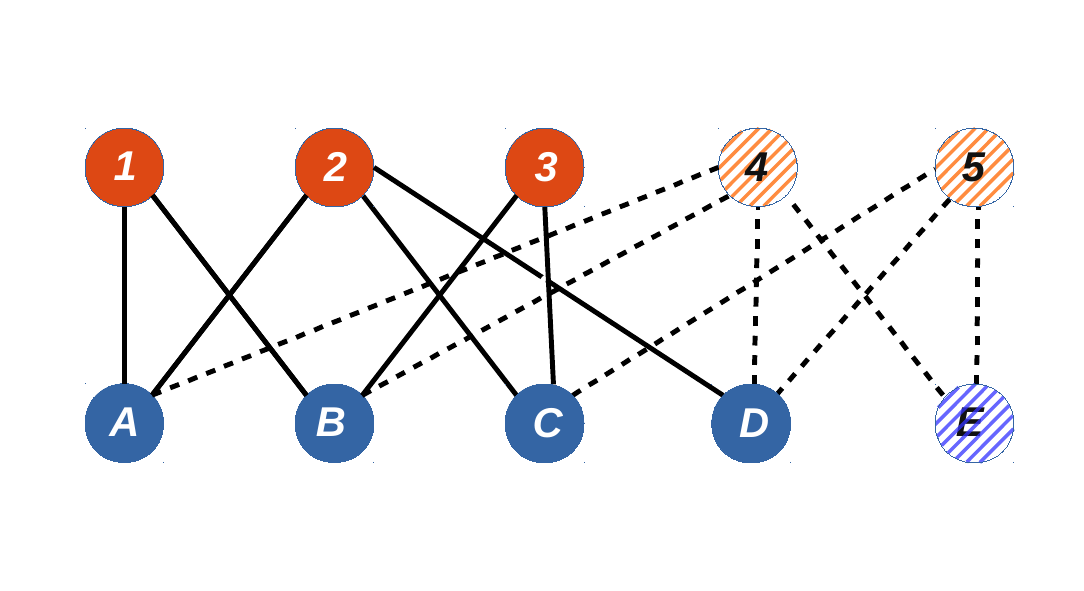}
	\caption{Schematic of how connections creating new four-cycles change the cycle basis of the graph and hide cycles of higher order. The initial cycle basis of the toy graph has one six-cycle only, $C=\{(1\textrm{B}3\textrm{C}2\textrm{A})\}$. With node 4 connected to A, B, D and E; and node 5 connected to C, D and E, the new cycle basis is the set $C=\{(2\textrm{D}4\textrm{A}),(4\textrm{D}5\textrm{E}),(2\textrm{C}5\textrm{D}),(1\textrm{B}4\textrm{A}),(3\textrm{C}5\textrm{D}4\textrm{B})\}$. Although we have two six-cycles in the network, (1\textrm{B}3\textrm{C}2\textrm{A}) and (3\textrm{C}5\textrm{D}4\textrm{B}), just the latter appears in the cycle basis.}
	\label{fig:4hiding6}
\end{figure*}

The effect of the increase of four-cycles in the network is twofold. First, a larger number of four-cycles means an increase in recurrent interactions between pairs of nodes, shifting the link weight distribution to the right (Figures \ref{fig:hgb_T3} and \ref{fig:hgb_T4}). Second, clustering in the projected network is stronger, even though the number of six-cycles in the cycle basis is smaller than in the case with higher temperature. That is, the change in temperature drastically changes the cycle basis of the network and the wider gap between the cycles, as mentioned above, hides the increase in the actual number of six-cycles in the network. The cycle basis is the set of cycles from which combinations can be made to create all other cycles in the graph. An example of how four-cycles can hide six-cycles in the network is shown in Figure \ref{fig:4hiding6}. We do not count the exact number of cycles of each size because counting cycles in a graph is a \textit{NP-complete} problem \cite{safar2014approximate}, i.e. it cannot be solved in polynomial time, and it is computationally too expensive to do so in large networks like ours.

Our proposed model does present a limitation that has yet to be overcome: none of the parameters of the model ($\alpha$, $T$, and $m$) seem to control degree-assortativity of the projected networks. Every bipartite network we built resulted in neutral degree-assortative projections. This is true even when we chose a heavy-tailed probability distribution to pick values of $m$ (not shown here), which turns the top degree distribution more right-skewed. The random characteristic of the model cannot capture the social factors driving assortativity in real-world projected networks.

However, the popularity vs. similarity model still represents well one-mode networks that are projections of a bipartite structure, otherwise as stated by its own creators in Section III C of \cite{papadopoulos2012popularity}. The bipartite version of the model can replicate the original bipartite network structures, such as top and bottom degree distributions and small cycles. As a result, the expected structural properties of projected networks (degree, clustering, and link weight distributions) naturally arise as part of the projection. We also notice that to model only the projection can be a misleading process. Instead, one should take a step back and consider modelling the bipartite network first, and only then, create its projection. 

\section{\label{sec:conclusion5}Conclusion}

In this work, we have introduced a generative model for bipartite networks, in order to better understand their structural properties. It is imperative that projections created using bipartite networks assessed with such a model can display features such as heavy-tailed degree and link weight distributions, and the high level of clustering, that are present in real one-mode networks.
 


By extending and adapting the popularity vs. similarity model proposed in \cite{papadopoulos2012popularity} to bipartite networks, we can control top degree distributions with a simple choice of a probability distribution. On the other hand, the tail of the bottom degree distribution is tuned by the parameter $\alpha$ of the model, ranging from peaked Poisson-like distributions to heavy-tailed power-law distributions. The frequency of the presence of small cycles can be tuned by controlling the temperature $T$ of the system. Therefore, we can recover degree distributions and the frequency of small cycles found in empirical bipartite networks.

Then, the structural properties of projected networks are straightforwardly inferred by building the projection out of the modelled bipartite network, except degree-assortativity. With both degree distributions of the bipartite network and the frequency of four-cycles, we naturally assess the resulting degree and link weight distributions of projections as found in real one-mode networks. The same is true for the clustering coefficients. Due to the high frequency of six-cycles in the network, the level of clustering in the projected one-mode network is not only the minimum level due to top degree nodes, as we discussed above. 

Finally, in contrast to the claim by the creators of the popularity vs. similarity model for one-mode networks \cite{papadopoulos2012popularity} that the model does not represent certain types of collaboration networks well, we have shown otherwise. The type of networks they referred to are actually one-mode projections of bipartite networks. We strengthened our claim, first presented in \cite{vasques2019bipartite}, that such projections should not be modeled directly, without taking in account the underlying bipartite structure. One should always consider the original bipartite network to assess the properties of networks in such cases.

%
%

\end{document}